%% file: IEEE-conference-template-062824.tex
\documentclass[conference]{IEEEtran}
\IEEEoverridecommandlockouts
\usepackage{amsmath,amssymb,amsfonts}
\usepackage{graphicx}
\usepackage{textcomp}
\usepackage{xcolor}
\usepackage{soul}
\usepackage{subcaption}
 \usepackage{amssymb}
\usepackage{mathtools}
\usepackage{textcomp}
\usepackage{stfloats}
\usepackage{xcolor}
\usepackage{booktabs} 
\usepackage{multirow}
\usepackage{url}
\usepackage{paralist}
\usepackage[noend]{algpseudocode}
\usepackage{balance}
\usepackage[ruled]{algorithm2e}
\usepackage{bbding}
\usepackage{pifont}
\usepackage[inline]{enumitem}
\usepackage{wrapfig}
\usepackage[noabbrev]{cleveref}
\crefname{section}{}{\S\S}
\crefdefaultlabelformat{\S#2#1#3}

\newcommand{\sys}{Easz\xspace}

\def\BibTeX{{\rm B\kern-.05em{\sc i\kern-.025em b}\kern-.08em
    T\kern-.1667em\lower.7ex\hbox{E}\kern-.125emX}}
\begin{document}

\title{Easz: An Agile Transformer-based Image Compression Framework for Resource-constrained IoTs
}

% \author{\IEEEauthorblockN{1\textsuperscript{st} Given Name Surname}
% \IEEEauthorblockA{\textit{dept. name of organization (of Aff.)} \\
% \textit{name of organization (of Aff.)}\\
% City, Country \\
% email address or ORCID}
% \and
% \IEEEauthorblockN{2\textsuperscript{nd} Given Name Surname}
% \IEEEauthorblockA{\textit{dept. name of organization (of Aff.)} \\
% \textit{name of organization (of Aff.)}\\
% City, Country \\
% email address or ORCID}
% \and
% \IEEEauthorblockN{3\textsuperscript{rd} Given Name Surname}
% \IEEEauthorblockA{\textit{dept. name of organization (of Aff.)} \\
% \textit{name of organization (of Aff.)}\\
% City, Country \\
% email address or ORCID}
% \and
% \IEEEauthorblockN{4\textsuperscript{th} Given Name Surname}
% \IEEEauthorblockA{\textit{dept. name of organization (of Aff.)} \\
% \textit{name of organization (of Aff.)}\\
% City, Country \\
% email address or ORCID}
% \and
% \IEEEauthorblockN{5\textsuperscript{th} Given Name Surname}
% \IEEEauthorblockA{\textit{dept. name of organization (of Aff.)} \\
% \textit{name of organization (of Aff.)}\\
% City, Country \\
% email address or ORCID}
% \and
% \IEEEauthorblockN{6\textsuperscript{th} Given Name Surname}
% \IEEEauthorblockA{\textit{dept. name of organization (of Aff.)} \\
% \textit{name of organization (of Aff.)}\\
% City, Country \\
% email address or ORCID}
% }
\author{
  Yu Mao\IEEEauthorrefmark{1},
  Jingzong Li\IEEEauthorrefmark{2},
  Jun Wang\IEEEauthorrefmark{3},
  Hong Xu\IEEEauthorrefmark{4},
  Tei-Wei Kuo\IEEEauthorrefmark{5},
  Nan Guan\IEEEauthorrefmark{3},
  Chun Jason Xue\IEEEauthorrefmark{1}
  \\
  \IEEEauthorrefmark{1}Mohamed bin Zayed University of Artificial Intelligence, UAE,
  \{yu.mao, jason.xue\}@mbzuai.ac.ae \\
  \IEEEauthorrefmark{2}The Hang Seng University of Hong Kong, Hong Kong, jingzongli@hsu.edu.hk \\
  \IEEEauthorrefmark{3}City University of Hong Kong, Hong Kong, \{jwang699-c, nanguan\}@my.cityu.edu.hk \\
  \IEEEauthorrefmark{4}The Chinese University of Hong Kong, Hong Kong, hongxu@cuhk.edu.hk \\
  \IEEEauthorrefmark{5}National Taiwan University, Taiwan, ktw@csie.ntu.edu.tw
}

\maketitle

\begin{abstract}
Neural image compression, necessary in various machine-to-machine communication scenarios, suffers from its heavy encode-decode structures and inflexibility in switching between different compression levels. 
Consequently, it raises significant challenges in applying the neural image compression to edge devices that are developed for powerful servers with high computational and storage capacities.
% The primary issue is that the computational and storage capabilities of edge devices are weaker than those of servers, preventing them from handling the same amount of computation and storage. 
%
%One solution is to downsample images and reconstruct them on the receiver side; however, current methods uniformly downsample the image and limit flexibility in compression levels. 
%
%In this work, we conduct the first investigation in neural image compression at edge devices, and propose
We take a step to solve the challenges by proposing 
a new transformer-based edge-compute-free image coding framework called 
\sys. \sys shifts the computational overhead to the server, and hence avoids the heavy encoding and model switching overhead on the edge. 
%
%\sys adopts a patch-erase algorithm to remove selected image contents with a conditional uniform-based sampler and then constructs the pixel using a transformer-based reconstruction on the receiver.
\sys utilizes a patch-erase algorithm to selectively remove image contents using a conditional uniform-based sampler. The erased pixels are reconstructed on the receiver side through a transformer-based framework.
%
% While \sys allows for flexible image size reduction, the computation on the receiver side becomes extremely unaffordable, even for a powerful server.
%
To further reduce the computational overhead on the receiver,
we then introduce a lightweight transformer-based reconstruction structure to reduce the reconstruction load on the receiver side. 
Extensive evaluations conducted on a real-world testbed demonstrate multiple advantages of \sys over existing compression approaches, in terms of adaptability to different compression levels, computational efficiency, and image reconstruction quality.
\end{abstract}


\input{sections/intro}

\input{sections/related}

\input{sections/design}
\input{sections/evaluation}

\input{sections/conclusion}

\bibliographystyle{ieeetr}
\bibliography{main}
% \printbibliography
\end{document}

%% file: sections/intro.tex
\section{Introduction}

%IoT scenarios can hardly benefit from the advancement of neural image compression.
%Neural-based image compression is facing a great challenge in its most important IoT scenarios. 
The need for advanced lossy image compression is raised by the explosive development of edge devices equipped with high-resolution cameras, such as industrial-inspections system~\cite{George2019}, wildlife observation system~\cite{wildlife}, and autonomous driving~\cite{zhang2023research},
%, i.e., street pedestrian detection system
%and unmanned aerial vehicle (UAV) fire-watch system,
by analyzing massive data sensed by heterogeneously connected devices.
Transmitting the huge volume of data generated by IoT devices can cause significant channel congestion, particularly in scenarios with limited bandwidth availability~\cite{huang2022compressive}. 
Neural-network (NN) based compressor can provide a better compression performance and outperform traditional image compression techniques like JPEG \cite{jpeg} and BPG \cite{bpg}. However, due to its heavy, symmetric encoding and decoding structure and inflexible compression rate adjustment, current NN-based methods have not yielded practical use on resource-constrained edge devices.

\begin{figure}[t]
\centering
\begin{subfigure}{0.48\textwidth}
    \includegraphics[width=\textwidth]{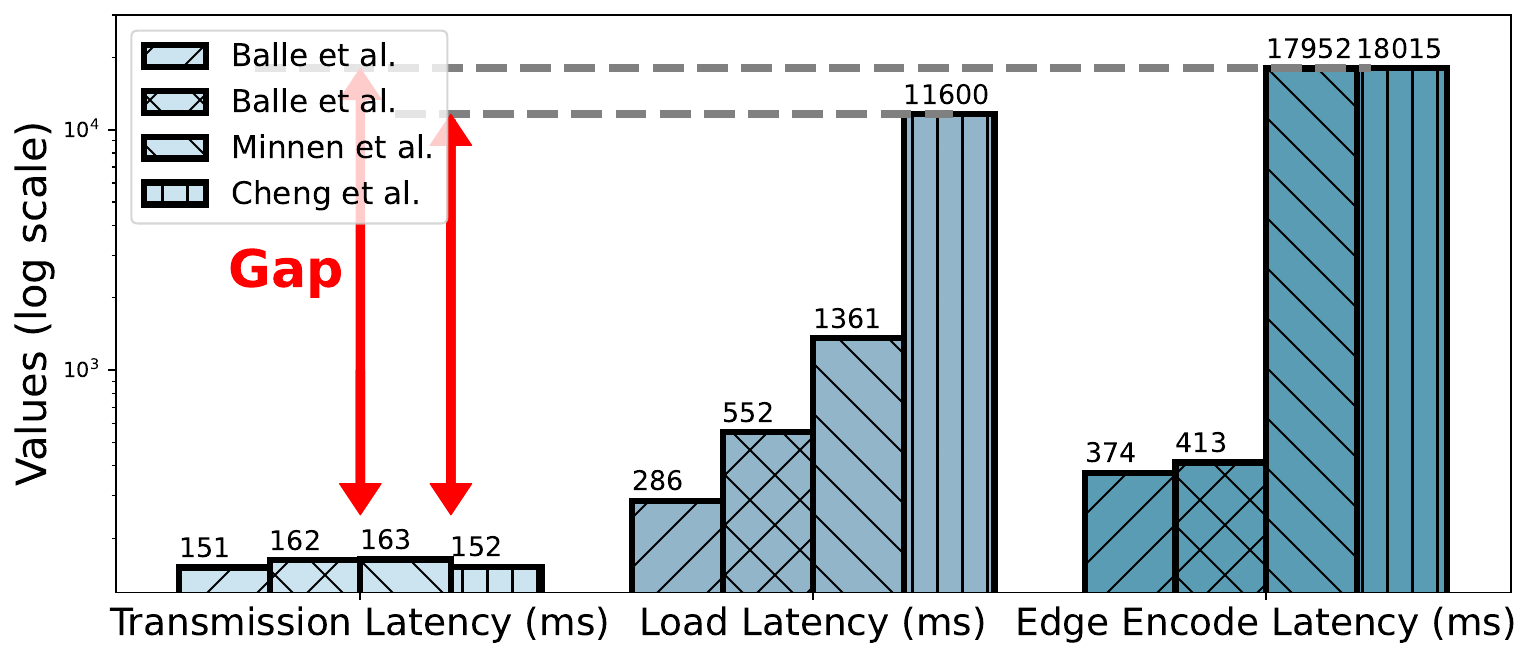}
\end{subfigure}
% \begin{subfigure}{0.24\textwidth}
%     \includegraphics[width=\textwidth]{pics/motivation.pdf}
%     \caption{}
%     \label{fig:quality-latency}
% \end{subfigure}
\caption{NN-based compressors face challenges on edge devices like the Jetson TX2, where loading and encoding an image can take over 10 seconds compared to a transmission latency of about 0.1 seconds.}
\label{fig:squeeze_exp}  
\end{figure}

Current edge NN-compression latency is unsatisfied due to the paucity of computational ability and storage on edge devices~\cite{FutureOfComputing, openai-ai-and-compute, moby, polly}.
%
%Given the paucity of computational ability and storage on edge devices in general~\cite{FutureOfComputing, openai-ai-and-compute, aws-outposts, moby, polly}, a huge gap would exist in the edge compression/decompress and transmission latency.
%
%The main challenge in neural image encoding on edge devices is the high computational demand, which far exceeds the processing capabilities of these devices~\cite{FutureOfComputing, openai-ai-and-compute, aws-outposts, moby, polly}, causing a huge gap between the transmission latency and edge execution latency. 
%
%This discrepancy results in a substantial gap between transmission latency and edge execution latency. 
%
As shown in Fig.~\ref{fig:squeeze_exp}, encoding a small $512\times768$ image can take as long as $18015$ ms on high-end devices equipped with a GPU like the NVIDIA Jetson TX2, not to mention that cameras usually capture images at 4K size.
%
%and even longer on less powerful devices such as the Raspberry Pi. 
Switching compression level can also cause $286\sim11600$ ms overhead, far exceeding the transmission time $151\sim163$ ms. %Therefore, the current NN-based compressor would definitely block the transmission and 

%Motivated by recent success in image generation, which has enabled several research fields including image super-resolution~\cite{choi2023n, chen2023activating}, image in-painting~\cite{guo2021image, chang2022maskgit} and image feature learning~\cite{he2022masked, tong2022videomae}, 

In this paper, we introduce \sys, an agile image coding framework that operates efficiently at both the edge and server side.
%
% \sys is compatible with all existing compression algorithms. 
%
%The intuition of \sys is an implicit assumption undermined in current solutions: the image needs to be uniformly downsampled.
%
To achieve agile encoding, \sys includes an erase-and-squeeze process, which removes the image patches based on a conditional random-based sampler. This technique can provide fine-grained compression level choices with fast adaptation to various scenarios. 
%
%The proposed conditional random-based sampler ensures flexibility for \sys. 
%However, this flexible encoding algorithm cannot be equipped with current reconstruction techniques, such as super-resolutions (whose implicit assumption is that the image is uniformly downsampled).
%
%However, it loses the chance to employ efficient reconstruction through convolution or the fast Fourier transform techniques.
%
To improve the reconstruction quality, we introduce the novel transformer architecture to conduct fine-grained pixel-level reconstruction.
%Under this setting, the reconstruction process has to be pixel-based, which is 
However, the pixel-level reconstruction using naive transformer architecture is computationally intensive ($ O(65536^2 \times 768)$ for a $256\times256$ image), even for a powerful 2080 Nvidia GPU server.
%(a 256 × 256 grayscale image would require 4, 294, 967, 296 × dmodel calculations).
%
We then propose a lightweight transformer architecture for efficient, high-quality reconstruction of erased patches on the receiver side. 
This involves a two-stage image patchify process to limit the scope of attention correlation calculations and a four-layer light weight transformer model for pixel-level local image reconstruction. 

The key contributions of this work can be summarized as:
\begin{itemize}
\item Generalized Erase-and-Squeeze Process: A new paradigm is introduced that offers more refined and flexible image reduction ratios;
\item Receiver-Side Lightweight Transformer Architecture: A super-lightweight (8.7MB) transformer architecture is designed for efficient and high-quality reconstruction of erased patches;
\item Compatibility with Existing Algorithms: \sys is compatible with all existing image compression algorithms and can also function independently;
\item We conduct extensive experiments that verify the effectiveness of \sys.

\end{itemize}

%% file: sections/related.tex
\section{Related Work}

% \noindent\textbf{Neural Network based Compression.} 

Image compression is experiencing significant growth~\cite{xie2024gaussian, xie2024output, minnenbt18, cheng2020image, mao2023faster}.
%with advancements in end-to-end training, hyperprior structures, entropy models, and encoder-decoder improvements
%~\cite{he2021checkerboard, mentzer2020high, mao2023faster, zhu2022transform,wang2023EVC,yang2022lossy,qian2022entroformer, wang2019towards, minnenbt18, kim2022joint, he2022masked, mao2022accelerating, plehn2022data}. 
Notable developments include the introduction of auto-regressive components~\cite{minnenbt18}, Gaussian Mixture Models for probability estimation~\cite{cheng2020image}, and general-purpose lossless compressiors using lightweight neural networks~\cite{mao2023faster, mao2022accelerating, mao2022trace}. 

%Attention mechanisms have been incorporated through Informer~\cite{kim2022joint}.

%, while Transformers and Swin architectures are replacing CNNs in encoding/decoding tasks~\cite{zhu2022transform, he2021checkerboard}. 

%Despite progress in real-world applications, challenges like inflexibility on switching models or high latency on the edge persist. The first problem is most NN-based image compressors need to switch models when changing compression levels. One attempt is to develop a model with a multi-branch that can deal with multiple compression ratios. However, these methods still need to deploy the encoder on the edge and suffer from difficult training for the multi-branch structure. Another trend is to downsample images on the edge and utilize a super-resolution method to reconstruct images on the server. In this way, the computational burden on the edge is alleviated. However, directly applying super-resolution in this scenario leads to an inflexible downsize rate and would also affect the reconstruction performance. 
Despite progress, real-world applications still face challenges at the edge~\cite{liang2025ariadne, reidy2024hirise, huang2022compressive}. ~\cite{reidy2024hirise} leverages in-sensor compression and selective ROI prioritization to optimize high-resolution image processing for edge ML.  
Deep-learning-based compression methods take about 1$\sim$20 second per image (512 $\times$ 768) on NVIDIA Jetson TX2, and many real-life endpoints are less potent than the TX2 (considering Raspberry Pi 4) but still need to compress images. A primary issue is that most NN-based image compressors require a model switch when changing compression levels.
~\cite{huang2022compressive} uses semantic-driven compressive sensing to enhance image compression for resource-constrained IoT systems.
Another approach involves downsampling images at the edge and using super-resolution techniques to reconstruct them on the server~\cite{yin2023online}. These methods reduce computational load at the edge, but applying super-resolution directly in this context results in an inflexible downsizing rate and can degrade reconstruction performance~\cite{laroche2023deep}.

%% file: sections/design.tex
\section{System Design}
\label{section3}

\begin{figure*}[t!]
\centerline{\includegraphics[width=0.8\linewidth]{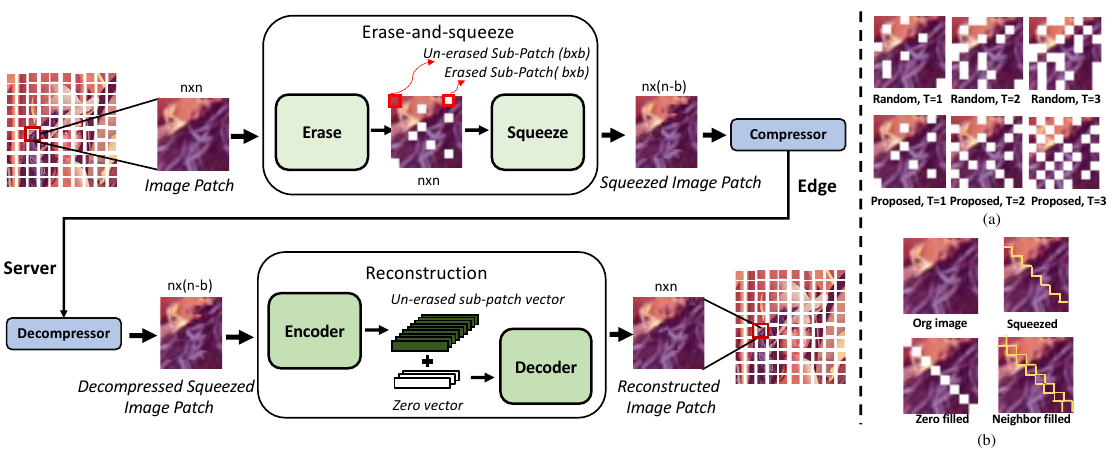}}
\caption{Left: \sys system overview. A image is going through a two-stage image patchify process, erase-and-squezee process. The squeezed image is then compressed using existing compressor and transmit. On the decompression stage, the erased patch is recovered.
Right: (a) Proposed erase methods compared with random erase methods. T indicates an erased item in each row. (b) Different methods to re-organize un-erased image components.}
\label{fig:main_idea}
\end{figure*}

% Our approach is based on the intuitive idea that an image contains redundant information, and neighboring pixels tend to exhibit robust local correlations. As a result, it becomes possible to predict missing patches using information from their neighboring patches. Guided by this intuition, we decided to erase some patches on the edge devices to improve compression rates, which comes with negligible additional computations. Then, a neural network is employed on the server side to predict the removed pixels. Following Fig.~\ref{fig:main_idea}, we introduce the system overview of \sys. 

The paper presents \sys, a novel edge-optimized image compression framework. It merges erase-based compression on the edge with a lightweight transformer-powered reconstruction on the server side, outperforming conventional codecs like JPEG and neural-network-driven compressors. 

% \subsection{System Overview}
%Our approach is based on the intuitive idea that selectively erases some patches on the edge devices to improve compression rates, which comes with negligible additional computations. Then, a neural network is employed on the server side to recover the removed patches. 

%Following Fig.~\ref{fig:main_idea}, we first introduce the system overview of \sys. 
%
%This involves splitting \(\widehat{S}\) into un-erased sub-patches \(\hat{U}\), which are repositioned using the same erasing mask \(M\). The position of erased sub-patch would be filled with zero value. These sub-patches are embedded in high-dimensional space before being encoded into high-dimensional representations by encoder \(E(\{0, \hat{U}\})\). A decoder \(D(E(\{0, \hat{U}\}))\) predicts the erased sub-patches \(\hat{U}'\) from un-erased features. Finally, transmitted and predicted sub-patches (\(\hat{U}\) and \(\hat{U}'\)) are combined to form the complete decompressed image \(\widehat{P}\).

\subsection{Erase-and-Squeeze Algorithm}
\label{sec:erase}
The section outlines the proposed \textit{Erase-and-Squeeze} strategy, including erase mask generation and the organization of un-erased image components on the edge/sender side.

% The Erase-and-Squeeze operation consists of three steps: 
% \begin{enumerate}
%     \item Divide the image patch into a grid of square sub-patches. The block size is determined by the user-specified sub-patch size $b$.
%     \item Erase $T$ sub-patches in each grid row, ensuring that the \textit{erased sub-patches} are not adjacent horizontally or vertically. The remaining block is referred to as \textit{un-erased sub-patches}.
%     \item Reassemble the un-erased patches into a complete image patch that can be compressed by other compressors.
% \end{enumerate} 

% Note that the first stage of image patchify, discussed in Sec.~\ref{sec:patchify}, is assumed to have been completed in this section. This allows us to concentrate on the erase-and-squeeze operation on the sub-patches.

% \begin{figure}[htbp]
% \centering
%         \includegraphics[width=0.8\linewidth]{pics/erase_block.pdf}
%         \caption{(a) Proposed erase methods comparing with random erase methods. T indicates erased item in each row. (b) Different methods to re-organize un-erased image components. Here we use a simple diagonal erasing mask for clear illustration.}
%         \label{fig:erase_block}
% \end{figure}

% \subsubsection{Erasing mask generation, rewrite to add ORB}
\textbf{Erase mask generation.} 
An erasing mask is a binary matrix used to determine if subpatches in an image patch should be kept or erased, defined as \(\mathbf{M}\) where each entry is set as follows:
\[
\mathbf{M}[i, j] = 1 \text{ if } (i, j) \text{ is sampled, else } 0
\]
%The mask ratio, which determines the proportion of the image that is sampled, is controlled by the patch size \( p \) and the sampled size \( T \). Specifically, the choice of patch size influences the granularity of the sampling, while the sampled size \( T \) dictates how many patches are included in the reconstruction for each row. By tuning these parameters, the framework can effectively balance the level of reconstruction difficulty and the computational load, enhancing model performance across various datasets. The mask will be sent with the compressed image for the receiver to decode.

%An erasing mask \(M\) is a binary matrix used to determine if subpatches in an image patch should be kept or erased. 
%
%a binary matrix of dimensions $(\frac{n}{b}, \frac{n}{b})$, is used to determine if subpatches in an image patch should be kept or erased. 
%If $M(i, j)$ is False, the subpatch at position $(i, j)$ is erased; if True, it's kept. 
Fig.~\ref{fig:main_idea} illustrates some mask examples with white blocks representing erased subpatches.
%In this paper both the server and the edge have access to this mask locally, although it doesn't cost much for transmission due to its tiny size – a binary 32x32 mask takes up just 128 bytes. 
%
As shown in Fig.~\ref{fig:main_idea}(b), a simple approach to erase is to remove blocks along the diagonal, which allows for easy reorganization of the remaining sections. However, the diagonal mask is not generalized due to its fixed erase ratio. Super-resolution-based approaches also suffer from a fixed file size reduction ratio due to uniform down-sampling.

\begin{figure}[t]
\centering
\begin{subfigure}{0.23\textwidth}
    \includegraphics[width=\textwidth]{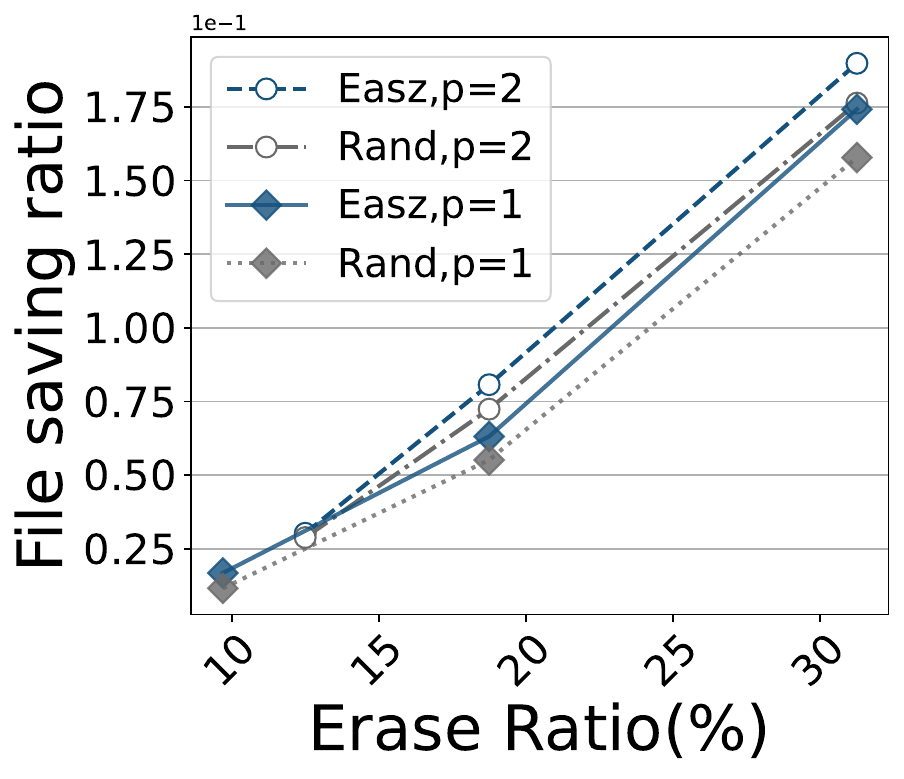}
    \caption{Impact on JPEG ($\uparrow$ better).}
    \label{fig:patch1_save_ratio}
\end{subfigure}
\begin{subfigure}{0.22\textwidth}
    \includegraphics[width=\textwidth]{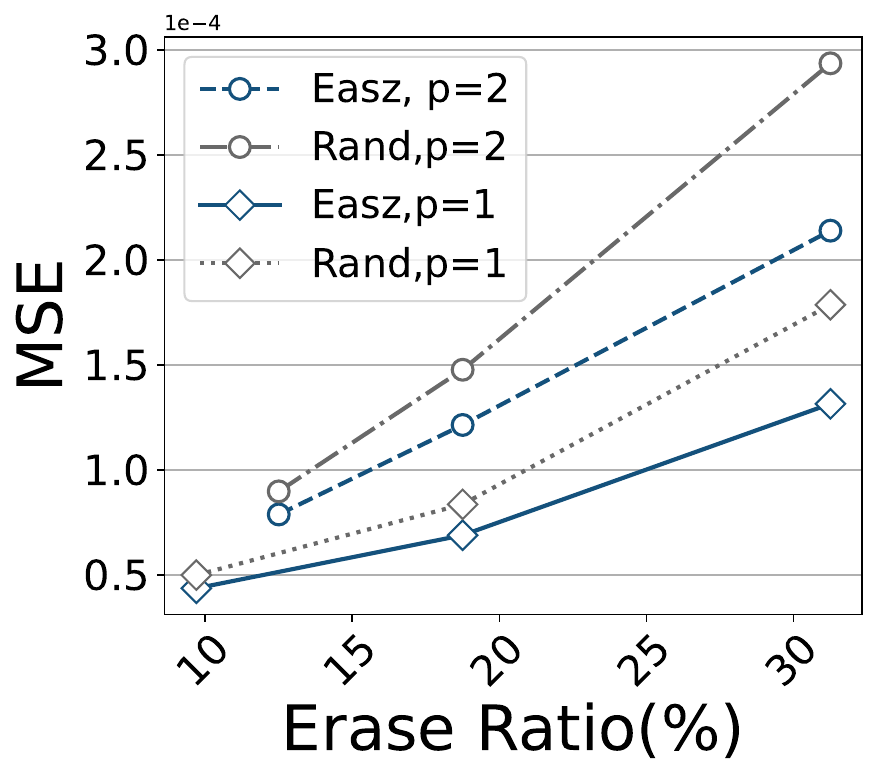}
    \caption{Impact on Recon ($\downarrow$ better).}
    \label{fig:patch1_psnr}
\end{subfigure}
\caption{The proposed method outperforms random masking in terms of JPEG impact and reconstruction, resulting in a higher file saving ratio and lower MSE on Kodak dataset. The variable $p$ represents patch size.}
\label{fig:sample_effect}
\end{figure}

To achieve flexible size reduction, a naive solution is to randomly erase a portion of patches. However, this can cause significant distortion due to consecutive information loss. Fig.~\ref{fig:main_idea}(a) shows a randomly generated mask that create large continuous erased areas, causing performance degradation in subsequent JPEG and reconstruction stages (See Fig.~\ref{fig:sample_effect}).

We then propose a generalized paradigm, a row-based conditional sampler for generating erase masks, covering various types including diagonal and uniform masks. This approach allows for a highly flexible sampling rate, while only need one model to reconstruct at any sampling rate. Also, it has better reconstruction behavior compared with a random masks.

\textbf{Row-based Sampler Definition}. 
%
%We first define a sampler \( G_r \) computes a mapping between the coordinates \( (i, j) \) of the downsampled image \( \hat{\mathbf{X}} \) and the relative coordinates \( (u, v) \) of the original image \( \mathbf{X} \). 
%
In this sampler, each row \( i \) of the image is processed sequentially, and within each row, the column coordinate \( j \) is selected using a sampler. Thus, the Row-based Sampler is defined by two functions, \( g_r^1 \), which governs the random column selection from \( \mathbf{X} \), while \( g_r^0(i) \) represents the current row:

\[
\hat{\mathbf{X}}[i, j] = \mathbf{X}[g_r^0(i), g_r^1(i, j)]
\]
where \( g_r^0(i) \) is a deterministic function representing row selection, and \( g_r^1(i, j) \) is a random mapping for the column coordinate within row \( i \).

%To ensure uniform random sampling within each row of the original image \( \mathbf{X} \), the row coordinate \( g_r^0(i) \) is fixed as \( i \), and the column selection function \( g_r^1(i, j) \) samples a pixel uniformly from the width \( W \) for each row \( i \).

\textbf{Constraints for Row-based Sampler.} When sampling from a matrix \(X \in \mathbb{R}^{H \times W}\), where each row is sampled \(T\) times, the new sample \(x_{i,t+1}\) is subject to:

1. Intra-row constraint (avoid proximity to previous samples in the same row):
% \resizebox{0.84\textwidth}{!}{
\begin{equation}
\begin{aligned}
    &g_r^1(i, t+1) \sim \text{Uniform}(0, W-1) 
    \\\quad 
    &\text{s.t.} \quad 
    \left| g_r^1(i, t+1) - g_r^1(i, t) \right| > \delta
\end{aligned}
\end{equation}
   % \[
   % g_r^1(i, t+1) \sim \text{Uniform}(0, W-1) \quad \text{subject to} \quad \left|g_r^1(i, t+1) - g_r^1(i, t)\right| > \delta
   % \]
   % }
   Here, \(\delta\) is a threshold distance that ensures the newly sampled column \(g_r^1(i, t+1)\) is sufficiently distant from the previously selected columns \(\{g_r^1(i, 0), \dots, g_r^1(i, t)\}\). 
   %This constraint guarantees a diverse selection of columns within each row, preventing the samples from clustering too closely together.

2. Inter-row constraint (minimize adjacency to prior samples from the preceding row):
\begin{equation*}
\begin{aligned}
   &g_r^1(i, t+1) \sim \text{Uniform}(0, W-1) \\ &\quad \text{s.t.} \quad \left|g_r^1(i, t+1) - g_r^1(i-1, T)\right| > \Delta
   \end{aligned}
\end{equation*}
   Similarly, \(\Delta\) represents a minimum separation between the newly sampled column in row \(i\) and the previously selected columns \(\{g_r^1(i-1, 0), \dots, g_r^1(i-1, T)\}\) from row \(i-1\). 
   %This prevents adjacent rows from sampling nearby columns, ensuring that the selection process avoids redundancy across rows.

Under these constraints, the row-based random sampler can be formalized as:
\vspace{-0.1cm}
\[
% \begin{aligned}
G_r = \left\{ 
    g_r^0(i) = I, 
    g_r^1(i, t+1)\right\},
\]
where
\[
\begin{aligned}
&g_r^1(i, t+1)\sim \text{clip}\left(\text{Uniform}(0, W-1)\right)\\
&\quad \text{s.t.} \quad 
        |g_r^1(i, t+1) - g_r^1(i, t)| > \delta \\
&\quad \quad\quad\ |g_r^1(i, t+1) - g_r^1(i-1, T)| > \Delta 
\end{aligned}
\]
% \vspace{-0.1cm}

%where the random column selection \(g_r^1(i, t+1)\) is adjusted dynamically to satisfy both the intra-row and inter-row constraints. This ensures a well-distributed sampling process across the entire matrix, balancing randomness with structured diversity.

%We first define the patch matrix $X$ with dimensions $n \times m$, and constrained that each row would be sampled $T$ times\footnote{This is to demonstrate the image could still be reassembled to a rectangle. We take row as example in this paper for clear explanation, but our method can also operate on column, too.}. The sampled results from each row should satisfy the following two conditions:1) In the same row, the new sample $x_{i,t+1}$ should avoid proximity to the previous $t$ samples as much as possible. 2) Across rows, the new sample $x_{i,t+1}$ should minimize adjacency to the prior $T$ samples $\{x_{i-1, 0}, \ldots, x_{i-1, T}\}$ from the preceding row. The sampling procedure is detailed in Algorithm~\ref{alg:sample} in Appendix. Diagonal masking and uniformly masking are subsets of the proposed algorithm. 

When restricted to T=1 with non-adjacent sampling in both row and columns, it becomes a diagonal mask. With patch=1, T=$n$/2, and non-adjacent sampling in each row and column, it degrades to 2x super-resolution. Also, we will show in experiment part that generating patches directly results in better reconstruction than existing super-resolution methods.

\textbf{Squeeze.} 
% After sampling, the downsampled image can be computed using this mask:
% \[
% \hat{\mathbf{X}}[i, j] = \begin{cases} 
% \mathbf{X}[g_g^0(i, j), g_g^1(i, j)] & \text{if } \mathbf{M}[i, j] = 1 \\
% 0 & \text{if } \mathbf{M}[i, j] = 0 
% \end{cases}
% \]
After applying the generated erase mask $\mathbf{M}$ on $\mathbf{X}$, the next step is to squeeze the non-zero (sampled) locations together to form a smaller image \({\mathbf{X}}_{\text{squeezed}} \in \mathbb{R}^{h' \times w' \times C}\), where \(h' < h\) and \(w' < w\) represent the dimensions. This can be achieved by filtering out the zero entries from \({\mathbf{X}}\):
\[
{\mathbf{X}}_{\text{squeezed}}[i', j'] = {\mathbf{X}}[i, j] \quad \text{for all } (i, j) \text{ where } \mathbf{M}[i, j] = 1
\]

After the process, the squeezed image \(\mathbf{X}_{\text{squeezed}}\) would be encoded using compressors like JPEG, BPG, etc to get a compressed form \(\hat{\mathbf{X}}_{\text{squeezed}}\) and send, as illustrated in Fig.~\ref{fig:main_idea}.

%\textbf{Discussion.} We emphasize again that under certain constraints, Easz's image removal strategy can degrade into a super-resolution image removal strategy. However, Easz is more comprehensive and offers more flexible options for image content removal. Easz is distinct from super-resolution because: 1) It does not require uniform downsampling of images. 2) Unlike super-resolution, which reconstructs low-level features from upsampled blurry images, Easz operates on partially empty image patches and predicts erased pixel values using surrounding data. 3) Easz can predict larger patches beyond single pixels. 
%Experimental results demonstrate that Easz surpasses traditional super-resolution in terms of PSNR and SSIM metrics when reducing pixel count equivalently. Figure~\ref{fig:detail} illustrates a comparison of image detail reconstruction, where Easz and other super-resolution all perform 2x reconstruction. It is evident that Easz better preserves image details; the children's faces are clearer, and the characters are more recognizable. In contrast, the super-resolution reconstructed image is unsatisfactory.

\begin{figure}[t]
\centering
        \includegraphics[width=\linewidth]{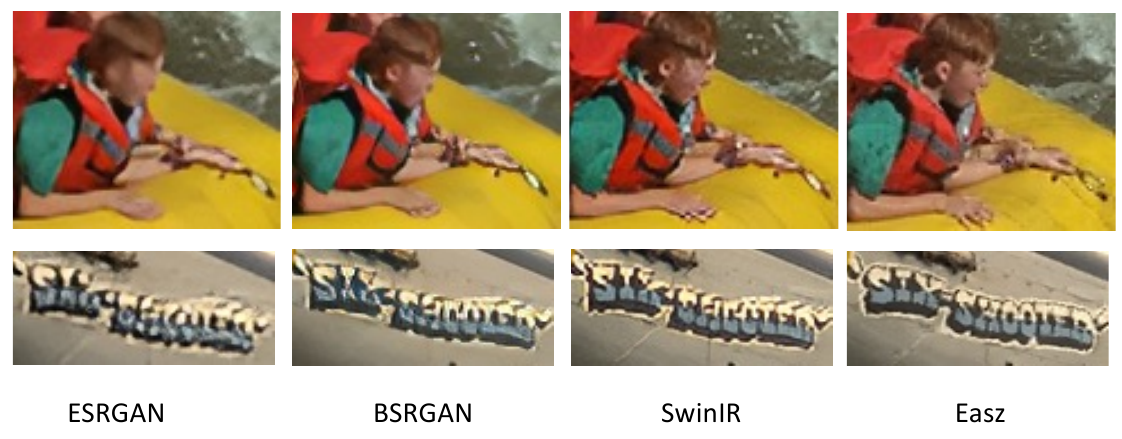}
        \caption{Easz preserves details better than SR methods via direct pixel prediction, improving PSNR and SSIM.}
        \label{fig:detail}
\end{figure}

\subsection{Reconstruction}
Recent work explores pixel-level generation for high-fidelity reconstruction~\cite{li2024mapping}. \sys adopts a transformer to perform pixel-level reconstruction, achieving better results than super-resolution methods, as shown in Fig.~\ref{fig:detail}. It also supports flexible downsampling ratios for adaptive compression.
However, directly apply transformers to predict pixels is costly due to the nature of attention mechanisms. 
In our experiments, a 12GB Nvidia 2080Ti server cannot afford to run pixel-based prediction of a single $256\times256$ image. Rather than applying conventional model compression techniques~\cite{mao2024compressibility}, our approach is inspired by the block-based strategies employed in traditional image compressors such as JPEG, which use $32 \times 32$ or $64 \times 64$ blocks to enhance compression performance. We design a two-stage image patchification process followed by a lightweight (8.4MB) transformer for sub-patch restoration.

\textbf{Complexity Analysis.} 
\label{sec:patchify}
Predicting pixel values for each element in the image, particularly in high-resolution images, proves to be computationally expensive. 
Consider a grayscale image of size 256x256 with a 1x1 patch size. This configuration results in 65,536 pixels. Based on the attention complexity, an image of size \((h, w)\) in grayscale would require $O((hw)^2\times d_{model})$ calculations if using a pixel as a token. Therefore, calculating a single self-attention for this setup would necessitate $65536^2 \times d_{model} = 4,294,967,296 \times d_{model}$ calculations. Such a computational load is heavy to execute on today's hardware, and it becomes exponentially more challenging when dealing with higher-resolution images.

\textbf{Two-Stage Image Patchify.} In the \sys framework, a two-stage image patchify process is employed to reduce the computational complexity of the vision transformer. Given an input image \(X\) with dimensions \((h, w)\), it is first divided into non-overlapping image patches of size \((n, n)\). Each of these \(n \times n\) image patches is further subdivided into smaller non-overlapping \((b, b)\) sub-patches. Critical operations such as erasure, squeezing, and reconstruction are performed at the sub-patch level. This patchify process is essential to manage the computational challenge of the vision transformer.

Next, we will explain how the proposed two-stage image patchify process reduces attention complexity. 
For clarity, the following analysis omit $d_{model}$ as it remains constant. 
The first stage patchify process split the image into $\frac{hw}{n^2}$ patches, with $n\times n$ size. 
%
%Since the complexity is determined by token numbers $\frac{hw}{n^2}$, if we stopped here and perform attention on this stage, 
%
The computational complexity on this stage is $O(\frac{(hw)^2}{n^4})$. 
%
%If we select $n>1$, some calculations are reduced. 
%
%However, as we'll analyze later, smaller token sizes lead to better compression performance. 
%
%In the extreme case where the token size is $1\times1$, we still aim to minimize model calculations. 
%
Then, the second patchify process is executed, split the $n\times n$ patches into $b\times b$ sub-patches. 
%
%By performing this, the token would be the sub-patches. 
%
Therefore there will be $\frac{(hw)^2}{n^2} \times \frac{n^2}{b^2}$ sub-patches, with $b \times b$ size. 
We configure the attention mechanism to operate within each image patch, thereby constraining the computational complexity to scale with the size of patches rather than the entire image, resulting in a computational complexity $O(\frac{hw}{n^2} \times \frac{n^4}{b^4} = O(\frac{(hw \times n^2)}{b^4})$. 
Therefore even when the token size is small, like when $b=1$, the calculation complexity is still $O(hwn^2)$, which is much smaller than the original complexity $O(hw)^2)$.

Now recall that image of size \((256, 256)\), which would cost $4,294,967,296 \times d_{model}$ calculations originally, after the two-stage image patchify process where $n=32$ and $b=4$, would require $1,048,576 \times d_{model}$ calculations, reduced by 4096 times compared to the original.

%By employing the two-stage patchify process (with \(n=32\) and \(b=1\)), this number is reduced by 256 times to $16,777,216 \times d_{model}$ calculations. The reduction in complexity comes from performing attention operations within each patch rather than across the whole image.

% \subsubsection{Two-Stage Image Patchifying}
% \label{sec:patchify}

% The \sys framework uses a two-stage image patchify process to reduce the computational complexity of vision transformers. An input image \(P\) with dimensions \((h, w)\) is first divided into non-overlapping patches of size \((n, n)\), which are then further split into smaller sub-patches of size \((b, b)\). This approach allows for critical operations at the sub-patch level and significantly reduces computational load. For example, a 256x256 grayscale image would require $4,294,967,296 \times d_{model}$ calculations if each pixel were treated as a token. By employing the two-stage patchify process (with \(n=32\) and \(b=1\)), this number is reduced by 256 times to $16,777,216 \times d_{model}$ calculations. Detailed analysis would be in Appendix. The reduction in complexity comes from performing attention operations within each patch rather than across the whole image.

\begin{figure}[t]
\centering
        \includegraphics[width=0.8\linewidth]{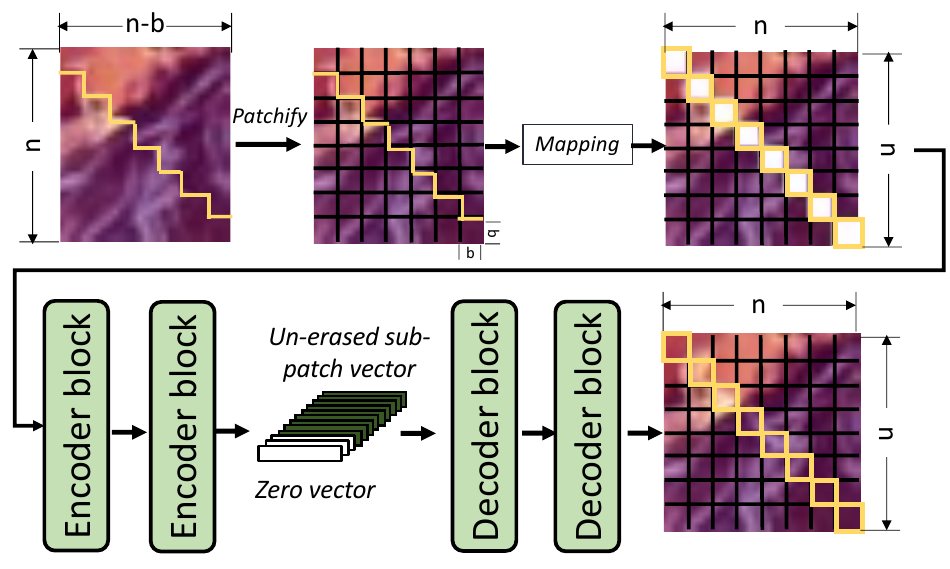}
        \caption{Reconstruct process illustration. }
        \label{fig:enc_dec}
\end{figure}

\textbf{Model Structure.}
%Reducing the size of a reconstruction not only lowers attention's computational demands but also results in a small overall model size. 
Fig.~\ref{fig:enc_dec} illustrates the architecture of our efficient transformer-based reconstruction network. The encoder and decoder are composed of two transformer blocks, each containing three layernorms, one attention layer, and one feedforward layer. 
% The formula for the described block is:
%
% \begin{align}
% U&=\text{layernorm2}(U+\text{Attention(layernorm1}(U))) \\
% U&=\text{layernorm3}(U+\text{FeedForward}(U))
% \end{align}
%
Note that the model can work with a variety of
input erase ratios, which are controlled by $k$ and $b$, and hence, we do not need to train a
specific model for each erase ratio. Sub-patches can execute in parallel for \sys due to the nature of transformer block calculations~\cite{vaswani2017attention}.

\textbf{Training Process.}  Consider a batch of image patches \(X_1, X_2, \ldots, X_n\) (each \(n \times n\)) randomly chosen from our training dataset, with patchification already completed. These patches are divided into smaller sub-patches of size (\(b \times b\)), resulting in (\(n/b, n/b\)) sub-patches per original patch. Positional embeddings are multiplied with these sub-patches to provide spatial context. An erase mask is then applied to remove \(k\) random sub-patches. The encoder \(E\) receives the remaining un-erased sub-patches $U = \{u_1, u_2, \ldots, u_m\}$ where $m=n/b \times n/b - k$. Each un-erased sub-patch $u_i$ is projected into an embedding $e_i$. The two-layer encoder processes these embeddings to extract features: $F = E(e_1, e_2, \ldots, e_m)$. To compensate for erased blocks during reconstruction, we introduce zero vectors $\hat{F} = \{\hat{f}_1,\hat{f}_2,\ldots,\hat{f}_k\}$. We align $\hat{F}$ with the zero-value positions and $F$ with one-value positions based on stored erase mask information. The combined feature sets $\{F,\hat{F}\}$ are fed into a two-layer decoder \(D\) which reconstructs the image: $\hat{X} = D(\{F,\hat { F }\})$. Our goal is to fine-tune encoders and decoders ($E$ and $D$, respectively) so that they minimize discrepancies between original images ($P$) and their reconstructions ($\hat {X}$), despite interference from zero vectors.

% \noindent\textbf{Edge Side:} The original image \(P_{h\times w}\) is divided into \(n \times n\) patches, which are further split into \(b \times b\) sub-patches. Using an erasing mask \(M\), each patch discards \(k\) sub-patches (denoted as erased sub-patches $U'$). The remaining sub-patches ($U$) are reorganized into a smaller rectangular patch. These patches are then reassembled to create the squeezed image \(S\), which is compressed with other compressor before transmission, denoted as middle compressor. The middle compressor used in this paper is JPEG due to its common use and prevalence. 

% \noindent\textbf{Server Side:} Upon receiving a compressed file, it is decompressed into \(\widehat{S}\) and then restored.
% %
% The process involves dividing \(\widehat{S}\) into un-erased sub-patches \(\hat{U}\), repositioning them using mask \(M\), and filling erased positions with zeros. These sub-patches are then embedded in high-dimensional space, encoded by encoder \(E(\{0, \hat{U}\})\), and decoded by \(D(E(\{0, \hat{U}\}))\) to predict the erased sub-patches \(\hat{U}'\). The final step combines transmitted and predicted sub-patches (\(\hat{U}\) and \(\hat{U}'\)) to create the complete decompressed image \(\widehat{P}\).

To minimize the difference between the original image $X$ and
the reconstructed image $\hat{X}$, we adopt LPIPS~\cite{zhang2018unreasonable}, a well-known perceptual loss, along with L1 loss as training loss. 
%$\lambda$ is chosen as 0.3 in our experiments.

\begin{align}
    \text{Loss}(x, y) &= \text{L1}(x, y) + \lambda*\text{LPIPS}(x, y)
    \label{eq:final_loss}
\end{align}
The feature extraction model is selected as VGG and $\lambda$ is chosen as 0.3 in our experiments.

%% file: sections/evaluation.tex
\section{Experiments}
\label{section4}

\subsection{Experimental Setting}

\noindent\textbf{Training setting.} The experiments consist of two phases: offline pretraining and online testing. In the pretraining phase, a specific loss function (Eq. \ref{eq:final_loss}) is used with these hyperparameters: learning rate of 2.8e-4, erase ratio of 0.25, batch size of 4096, weight decay of 0.05. Randomly generated erase masks are applied for model robustness during this stage. For online testing, a consistent mask is utilized on both edge and server sides. But note that transmission of the mask won't be costly due to its small size—a binary mask at dimensions $32 \times 32$ occupies only 128 bytes.

\noindent\textbf{Hardware platforms.}
Our framework is implemented with $\sim$1000 lines of Python. We use an NVIDIA Jetson TX2 as the edge device and a desktop with Intel i7-9700K CPU and RTX 2080Ti GPU as the server, which are physically connected to a Wi-Fi router and communicate via TCP. 

% \begin{equation}
%     LR = blr*min(\frac{e + 1}{we + 1^{-8}}, 0.5 * (cos(\frac{e}{te} * pi) + 1))
% \end{equation}

% in which blr is base learning rate, e is epoch, we is warm epoch, and te is total epoch.

\noindent\textbf{Datasets.} During the offline pretraining phase, the CIFAR-10~\cite{krizhevsky2009learning} dataset is employed to pretrain the model, enabling it to acquire generative capabilities. In the testing phase, two common image compression datasets, Kodak~\cite{kodak_dataset} and CLIC~\cite{clic_dataset}, are utilized to assess the generative performance of the proposed method.
% \begin{itemize}
%     \item Kodak Dataset~\cite{kodak_dataset}: This dataset comprises 24 lossless, true-color images, with each image having 24 bits per pixel (also known as "full color"). These images have been made available by the Eastman Kodak Company for unrestricted usage. The image dimensions in this dataset are 768x512 pixels.
%     \item CLIC Dataset~\cite{clic_dataset}: The CLIC dataset is derived from the CLIC 2022 test set and contains 30 images sourced from Unsplash. These images are provided in PNG format and have been resized such that the longer side is 2048 pixels.
% \end{itemize}

\noindent\textbf{Metrics.}  To benchmark against other super-resolution approaches, we include PSNR and SSIM to demonstrate the superiority of our method. Since removing content would generally impact reference-based metrics such as PSNR and SSIM (a trend also observed in other downsampled-and-super-resolution methods), we employ non-reference perceptual metrics for comparison with other compression baselines: Brisque~\cite{mittal2012no}, Pi~\cite{blau20182018}, and Tres~\cite{golestaneh2022no}.
Compression performance is evaluated by bits per pixel (BPP).

% \begin{table}[t!]
% \caption{The image quality metrics we adopt.}
% \begin{center}
% \begin{tabular}{|c|c|c|c|c|}
% \hline
% Metrics & Brisque & Pi & Tres & MSE\\
% \hline
% Higher Better? & \ding{56} & \ding{56} & \ding{52} & \ding{56}\\
% \hline
% \end{tabular}
% \label{tab:metric}
% \end{center}
% \end{table}

% PSNR is a mathematical measurement of the quality of the reconstruction of a compressed image. It measures the ratio between the maximum possible value of a signal and the noise that affects the signal. Higher values of PSNR indicate better image quality. 
% \begin{equation}
% PSNR = 10 \cdot \log_{10} \left( \frac{MAX^2}{MSE} \right)    
% \end{equation}
% where $MAX$ is the maximum possible pixel value of the image (usually 255 for 8-bit images), and $MSE$ is the mean squared error between the original and compressed images.

\noindent\textbf{Baselines.} We use four compression methods as baselines to demonstrate the effectiveness of the proposed method: JPEG, BPG, MBT\cite{minnenbt18}, and Cheng-Anchor~\cite{cheng2020image}. Among these, MBT and Cheng-Anchor are two NN-based compression methods. 

\subsection{Latency Analysis and Resource Consumption} 
\label{section4.2}

\begin{figure}[t]
% \begin{wrapfigure}{l}{0.5\textwidth}
%     \centering
\begin{subfigure}{0.47\textwidth}
    \includegraphics[width=\textwidth]{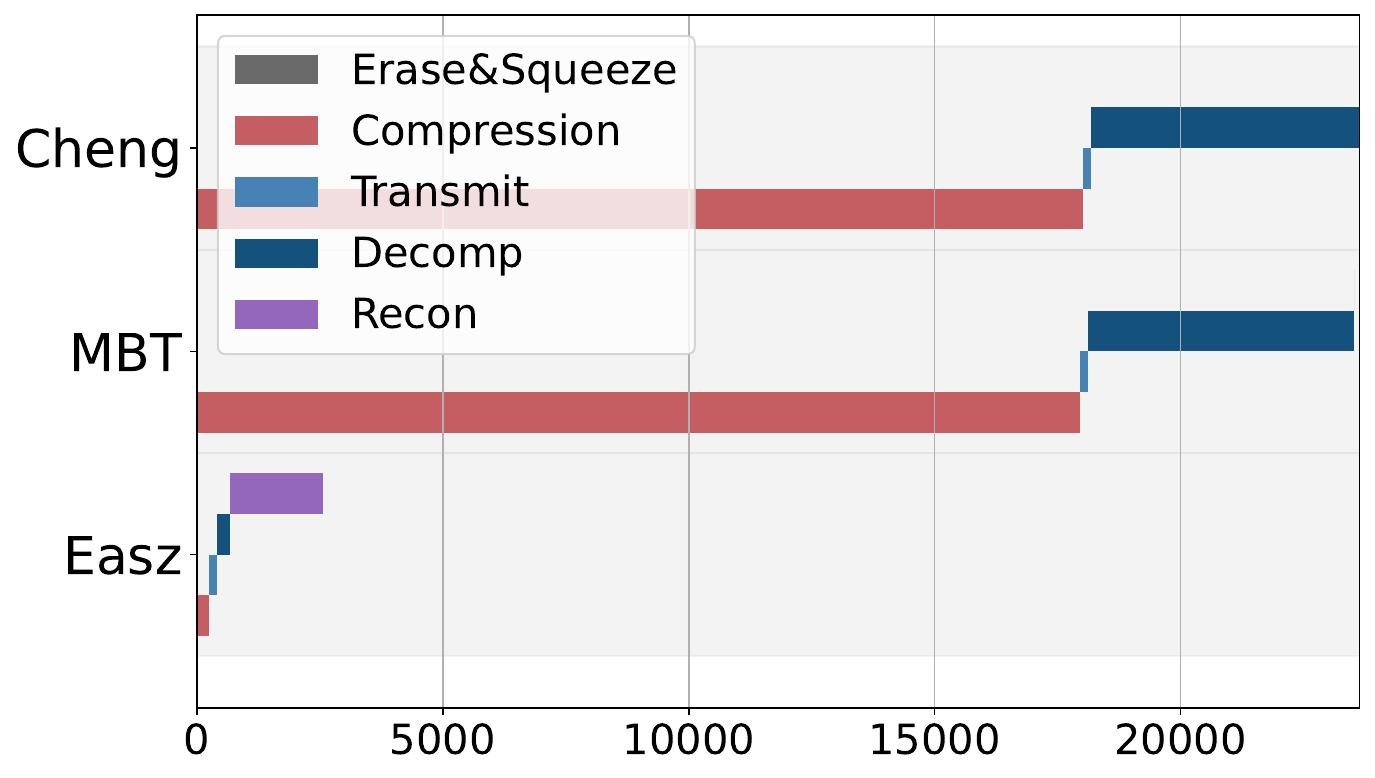}
    \caption{End-to-End Latency Breakdown(ms).}
    \label{fig:latency_break}
\end{subfigure}
\begin{subfigure}{0.24\textwidth}
    \includegraphics[width=\textwidth]{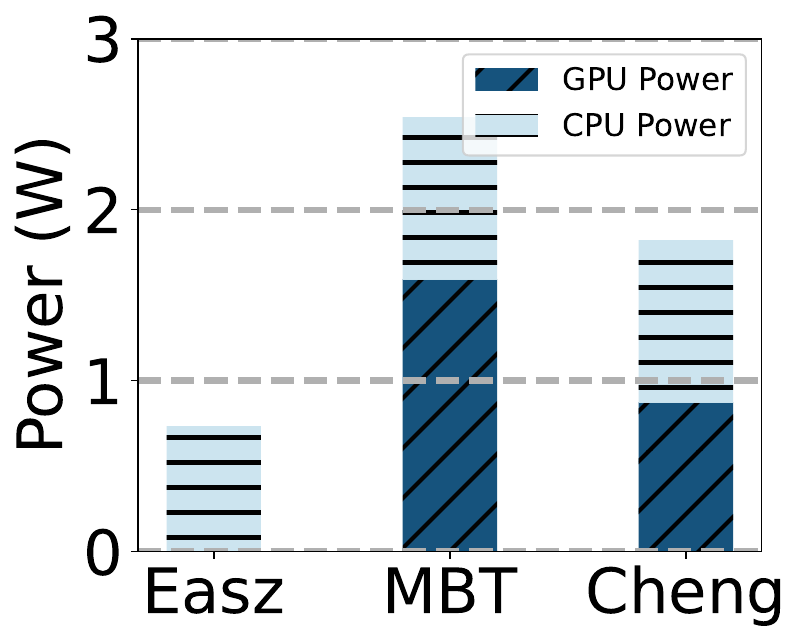}
    \caption{Encode Power consumption.}
    \label{fig:power}
\end{subfigure}
\begin{subfigure}{0.24\textwidth}
    \includegraphics[width=\textwidth]{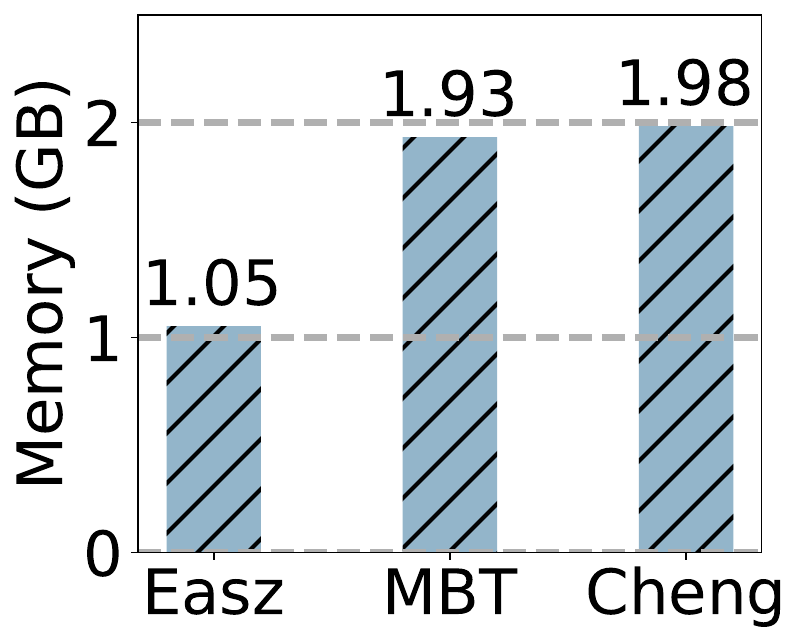}
    \caption{Encode Memory footprint.}
    \label{fig:memory}
\end{subfigure}
\caption{Efficiency Evaluation on NVIDIA Jetson TX2.}
\label{fig:resource}  
% \end{wrapfigure}
\end{figure}

We first report the latency breakdown of \sys with other neural network-based compression methods, using a Jetson TX2 for compression and a server for decompression. We repeat the runs 24 times and report the average on Fig.~\ref{fig:latency_break}. We observe that Erase-and-Squeeze only takes up 0.7\% of the end-to-end latency, which induces minimal overhead on the edge device and proves \sys' efficiency,
while both MBT and Cheng-Anchor are too compute-intensive to run the compression on the edge side. 
As expected, the reconstruction in \sys takes the longest time, accounting for 74\% of the latency.
We argue that the performance can be significantly improved by upgrading to a datacenter-class GPU such as the A100. Additionally, acceleration using FPGAs presents an interesting direction for future exploration~\cite{tang2022p,tang2024stem}.
%This improvement is due to its architecture designed for low overhead, offloading computational tasks to servers instead of edge devices, and using the JPEG codec for efficient operation with less resource consumption.

% \noindent\textbf{Latency breakdown.}
% To further analyze the execution time of each key component, we repeat the runs 24 times and report the average on Fig.~\ref{fig:latency_break}.
% Erase-and-Squeeze operation only takes up 0.7\% of the end-to-end latency, which induces minimal overhead on the edge device and proves the efficiency of \sys' design, while other state-of-the-arts are too compute-intensive to run the compression on the edge side. 
% As expected, the reconstruction in \sys takes the longest time, accounting for 74\% of the latency.
% We argue that it can be significantly improved by upgrading to a datacenter-class GPU, such as the A100, instead of the RTX 2080Ti.

Resource consumption is another critical consideration when it comes to resource-constrained edge devices. To assess this, we measure three key metrics – CPU power, GPU power, and memory footprint – using the Tegrastats Utility \cite{tegrastats} on the Jetson TX2. As illustrated in Fig.~\ref{fig:resource}, our findings reveal that \sys excels in all metrics compared to other NN-based compression methods. Specifically, in contrast to MBT and Cheng-Anchor, Easz achieves a remarkable 71.3\% and 59.9\% reduction in total power consumption. It's noteworthy that Easz does not utilize any GPU power on the edge device, attributed to its lightweight yet effective erase-and-squeeze design. Furthermore, \sys reduces memory footprint by 45.8\% and 47.1\%, respectively. These results underscore the advantage of deploying \sys on wimpy edge devices.

\subsection{Comparison with Super-Resolution Methods}
\label{section4.3}

We compare the reconstruction effect of Easz with state-of-the-art super-resolution methods to demonstrate Easz's effectiveness. As shown in Tab.~\ref{tab:super_res}, Easz outperforms super-resolution in pixel-level reconstruction metrics while having a much more flexible reduction ability. Note that Easz uses a model of only 8.7MB, while other models are 67MB.

\begin{table}[t]
  \centering
  %\small
  \setlength\tabcolsep{4pt}
  \caption{Comparison with Super-Resolution on Kodak Dataset.}
    \begin{tabular}{|c|c|c|c|c|}
    \hline
    Metrics & Easz &SwinIR  & realESRGAN  & BSRGAN   \\
    \hline
    PSNR & 28.96 & 24.86 & 24.85 &25.35\\
    MS\_SSIM & 0.96 & 0.94 & 0.93 & 0.94\\
    Recon Model Size & 8.7MB & 67MB & 67MB & 67MB\\
    \hline
    \end{tabular}%
  \label{tab:super_res}%
\end{table}%

% \subsection{Impact analysis on image analytic tasks}
% \label{exp:classification}

% Another aspect to compare is how different methods affect subsequent image analysis. We select image classification as example.

%Our method was tested for its impact on image classification accuracy with high-dimensional features, comparing JPEG and BPG compression methods using the Kodak dataset. Baseline accuracies were affected by the compression level: JPEG had a consistent 87.5\% accuracy across three levels, while BPG varied from 95\% to 66\% depending on quality settings (20 to 40). Incorporating our method showed that up to a certain erase ratio threshold, classification accuracy is maintained or improved; exceeding this threshold reduces accuracy. The critical erase ratios differed based on quality settings: for JPEG they were at 6.25\%, 18.75\%, and 25\%, and for BPG at 12.5\%, 18.75\%, and 25\%. 

\subsection{Ablation Study}
\begin{figure}[t]
\centering
\begin{subfigure}{0.2\textwidth}
    \includegraphics[width=\textwidth]{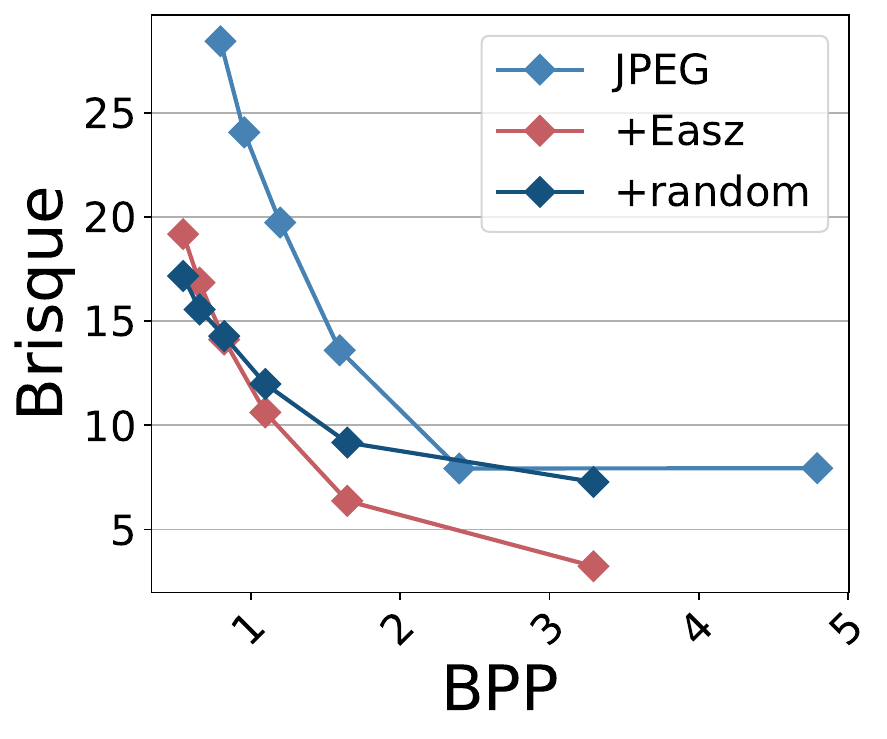}
    \caption{JPEG, Brisque($\downarrow$ better).}
    \label{fig:brisque_jpeg_ab}
\end{subfigure}
% \begin{subfigure}{0.3\textwidth}
%     \includegraphics[width=\textwidth]{pics/brisque_webp_ab.eps}
%     \caption{WebP, Brisque.}
%     \label{fig:brisque_webp_ab}
% \end{subfigure}
\begin{subfigure}{0.19\textwidth}
    \includegraphics[width=\textwidth]{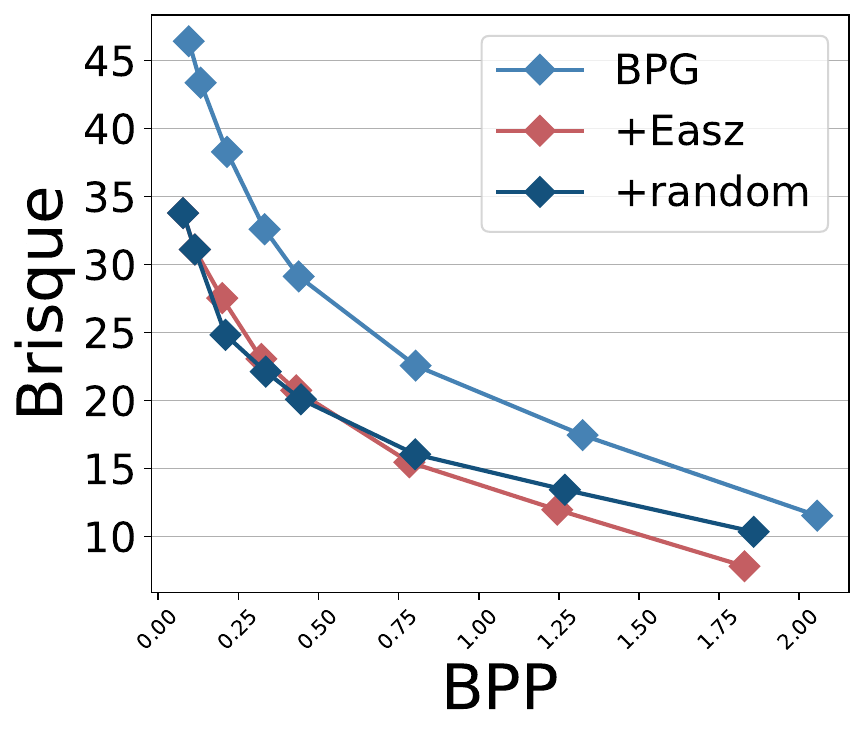}
    \caption{BPG, Brisque($\downarrow$ better).}
    \label{fig:brisque_bpg_ab}
\end{subfigure}
\begin{subfigure}{0.2\textwidth}
    \includegraphics[width=\textwidth]{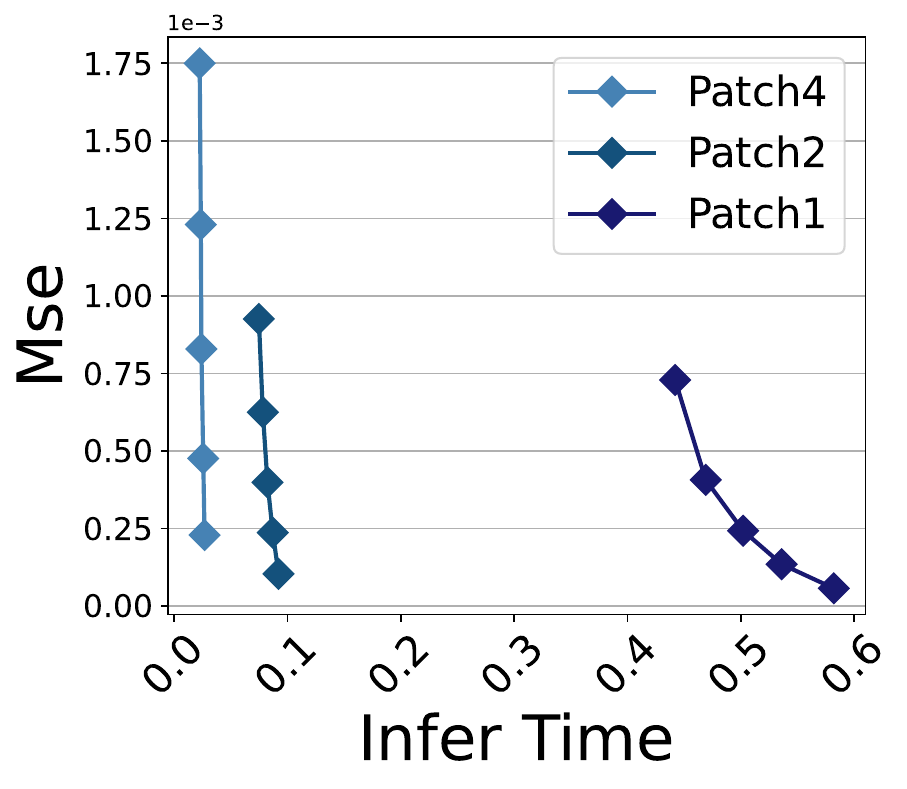}
    \caption{MSE($\downarrow$ better).}
    \label{fig:patch_selection}
\end{subfigure}
\begin{subfigure}{0.2\textwidth}
    \includegraphics[width=\textwidth]{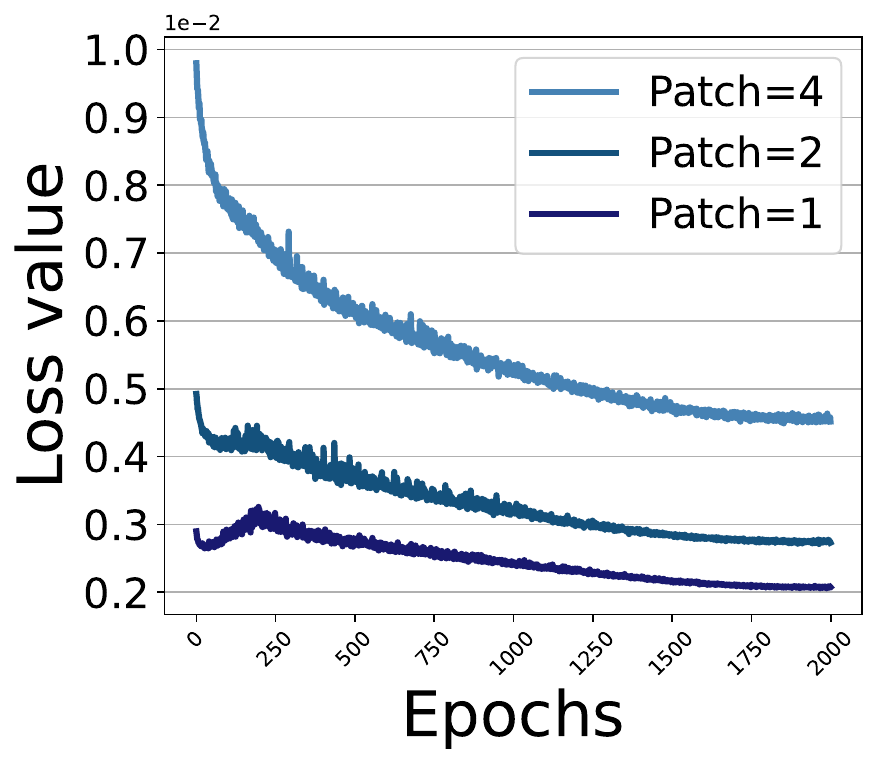}
    \caption{MSE($\downarrow$ better)}
    \label{fig:finetune_loss}
\end{subfigure}
\caption{(a)(b): Comparison between \sys with proposed mask strategy, \sys with random mask strategy, and conventional compression baselines (JPEG and BPG). (c) Patch size and erase ratio's impact on MSE. (d)  MSE during \sys fine-tuning process with patch size=1,2,4 on Kodak dataset.}
\label{fig:brisque_ab}  
\end{figure}

\textbf{Effectiveness of erase strategy.}
Fig.~\ref{fig:brisque_jpeg_ab} and Fig.~\ref{fig:brisque_bpg_ab} compare the proposed to erase mask generation method, the random erase mask method, and the baseline (JPEG and BPG) throughout the entire pipeline. It can be observed that the proposed erase mask generation method achieves better BPP at the same quality level on both JPEG and BPG, further substantiating the effectiveness of the proposed method.

\textbf{Patch size selection.}
Fig.~\ref{fig:patch_selection} examines the effects of two hyperparameters, erase block size (1, 2, and 4) and erase ratio (10\% to 50\%), on compression rate and quality. As the erase ratio increases, MSE rises, indicating lower reconstruction quality. Smaller patch sizes yield better reconstruction due to higher local correlations. Patch size=2 offers a balance between speed—being six times faster than size=1—and quality—with only a slight difference in MSE. Doubling the patch size from 2 to 4 also doubles both speed and MSE. The recommendation is to use smaller patch sizes for practical applications but consider size=2 for additional speed needs.

%Two hyperparameters can control the erase proportion: erase block size and erase ratio. We will consider erase block sizes of 1, 2, and 4 and erase ratio from 10\% to 50\%  to study the impact of these two hyperparameters on compression rate and compression quality.
%From Fig.~\ref{fig:patch_selection}, we can observe that the MSE increases as the erasure ratio increases, signifying a decline in reconstruction quality. The MSE goes down with the patch size, indicating better reconstruction quality. This result is expected because a smaller patch size implies that un-erased sub-patches are more likely to be closer to erased sub-patches, resulting in higher local correlations. When comparing patch size=2 with patch size=1, the difference in reconstruction quality is not substantial, with a maximum gap of only 10\% as shown in Fig.~\ref{fig:patch_selection}. However, the reconstruction speed is 6x faster. On the other hand, when comparing patch size=4 with patch size=2, the speed doubles, but the MSE also doubles. Therefore, we recommend using a smaller patch size whenever possible in practical applications. If further speed requirements exist, patch size 2 is a good choice.

\textbf{Effectiveness of fine-tuning.} 
Our model, after pretraining on the CIFAR dataset for 5000 epochs, can be applied to various image compression tasks due to its ability to recognize similarities in local image features. Typically, models are first pre-trained on a large dataset and then fine-tuned for specific tasks. We tested if fine-tuning our pretrained model with the Kodak dataset would be beneficial and found that it indeed improves performance by reducing losses across different patch sizes ($1 \times 1$, $2 \times 2$, and $4 \times 4$), as shown in Fig.~\ref{fig:finetune_loss}. 
%This suggests that online fine-tuning of pre-trained models could further enhance compression effectiveness in real-world applications.
%By leveraging the similarity between local features in images, our model can be generalized to other image compression tasks after pretraining for 5000 epochs on the CIFAR dataset. However, models typically undergo two phases in traditional visual tasks: pre-training and fine-tuning. Initially, the model is trained on a large dataset to learn a robust image representation. Later, during fine-tuning, it's adapted for optimal performance on a specific task dataset. It raises whether fine-tuning the target dataset would benefit \sys. We fine-tune the pre-trained model using the Kodak dataset, which consistently leads to further improvements by reducing losses for different patch sizes (1x1, 2x2, and 4x4), as depicted in Figure~\ref{fig:finetune_loss}. This demonstrates that fine-tuning deployed pre-trained models under practical scenarios can enhance compression performance. This observation opens up the possibility of online fine-tuning the model on the server.

% \subsubsection{Effectiveness of loss function}

\subsection{Improvement on Existing Compressors}

To evaluate how well \sys works with leading compressors, we incorporated it into four established methods: JPEG and BPG (traditional compressors), as well as MBT and Cheng-anchor (neural network-based compressors). We used two datasets, Kodak and CLIC, to test the resilience of \sys across different types of image data. For the Kodak dataset, we aimed for a bit-per-pixel (BPP) rate of approximately 0.4; for the CLIC dataset, we targeted a BPP of around 0.3 to gauge \sys's efficacy at varying levels of compression. The results showing how each baseline method performs on its own and when combined with \sys are detailed in Tables~\ref{tab:kodak}. The data clearly shows that integrating these baselines with \sys consistently enhances perceptual quality without increasing BPP. 

\begin{table}[t]
  \centering
  %\small
  \setlength\tabcolsep{0.5pt}
  \caption{Compression Performance Enhancement on Kodak Dataset and Clic Dataset.}
  \resizebox{0.5\textwidth}{!}{
    \begin{tabular}{|c|c|cc|cc|cc|cc|cc|}
    \hline
    \multicolumn{2}{|c|}{\multirow{2}{*}{Metrics}} & \multicolumn{2}{c|}{JPEG} & \multicolumn{2}{c|}{BPG} & \multicolumn{2}{c|}{MBT} & \multicolumn{2}{c|}{Cheng-anchor}\\
    \cline{3-7}  \cline{8-10}    \multicolumn{2}{|c|}{} & \textit{Org} & \textit{+Proposed} & \textit{Org} & \textit{+Proposed} & \textit{Org} & \textit{+Proposed} & \textit{Org} &  \textit{+Proposed}\\
    \hline
    & BPP  &  0.412 &  0.411 & 0.382  &  0.410  & 0.433 & 0.389 & 0.418 & 0.402\\
    \cline{2-10}
    \multirow{3}{*}{Kodak} & Brisque  & 43.06  &  22.34 &  30.675  & 23.27  & 28.13 & 18.63 & 29.16
 & 20.51\\
    & Pi  & 4.84  & 3.33  & 3.07  & 3.04  & 3.01 & 3.00 & 3.11
& 3.05 \\
    & Tres   & 77.62  & 86.26  & 83.55 & 85.88  & 84.14 & 88.03 & 88.53 & 89.80\\
    \hline
    & BPP  & 0.306 &  0.307 &  0.308 & 0.293  & 0.308 & 0.292 & 0.287 & 0.267\\
    \cline{2-10}
    \multirow{3}{*}{Clic} & Brisque  &  60.51 &  23.63 &  39.95 &  25.27 & 32.20 & 18.37 & 35.42 & 21.55\\
    & Pi  & 8.51  & 5.02  & 4.85  & 4.66  & 4.33 & 4.35 & 4.58 & 4.50 \\
    & Tres   &  50.65 &  63.69 &  65.14 &  67.08 & 73.54 & 78.30 & 82.91 & 83.95 \\
    \hline
    \end{tabular}%
  }
  \label{tab:kodak}%
\end{table}%

\subsection{End-to-End Compression Performance}

\begin{figure}[t]
\centering
\begin{subfigure}{0.2\textwidth}
    \includegraphics[width=\textwidth]{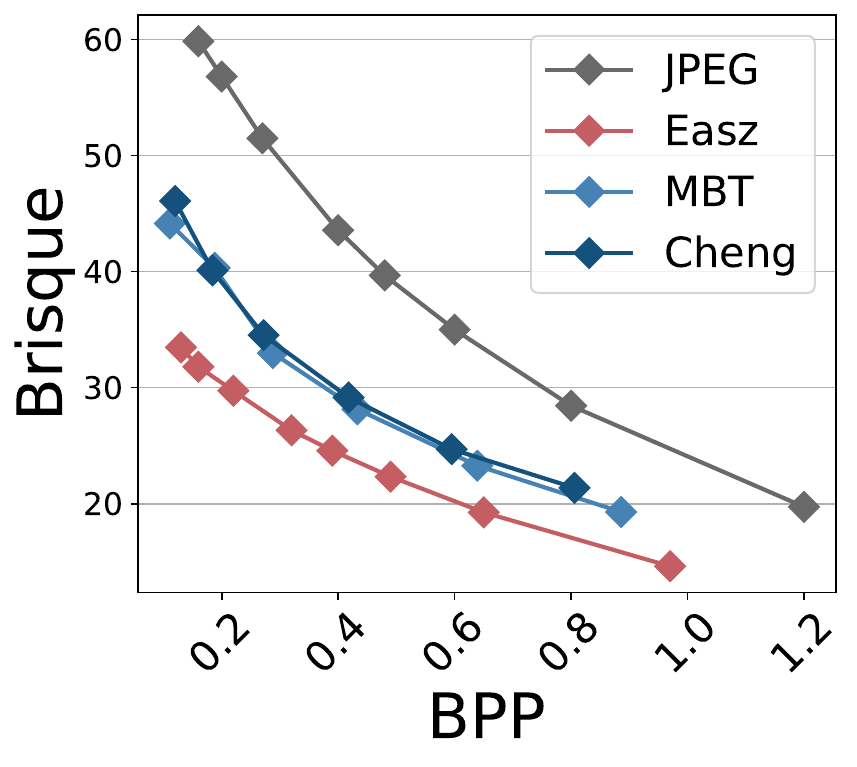}
    \caption{Brisque ($\downarrow$ better).}
    \label{fig:distortion_jpeg}
\end{subfigure}
\begin{subfigure}{0.19\textwidth}
    \includegraphics[width=\textwidth]{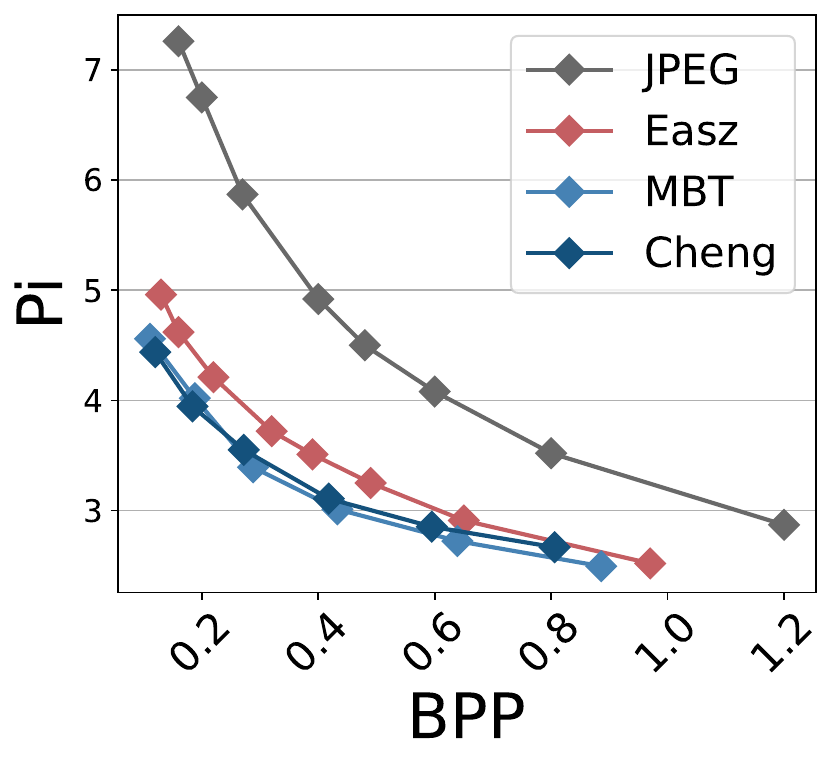}
    \caption{Pi ($\downarrow$ better).}
    \label{fig:distortion_bpg}
\end{subfigure}
\begin{subfigure}{0.2\textwidth}
    \includegraphics[width=\textwidth]{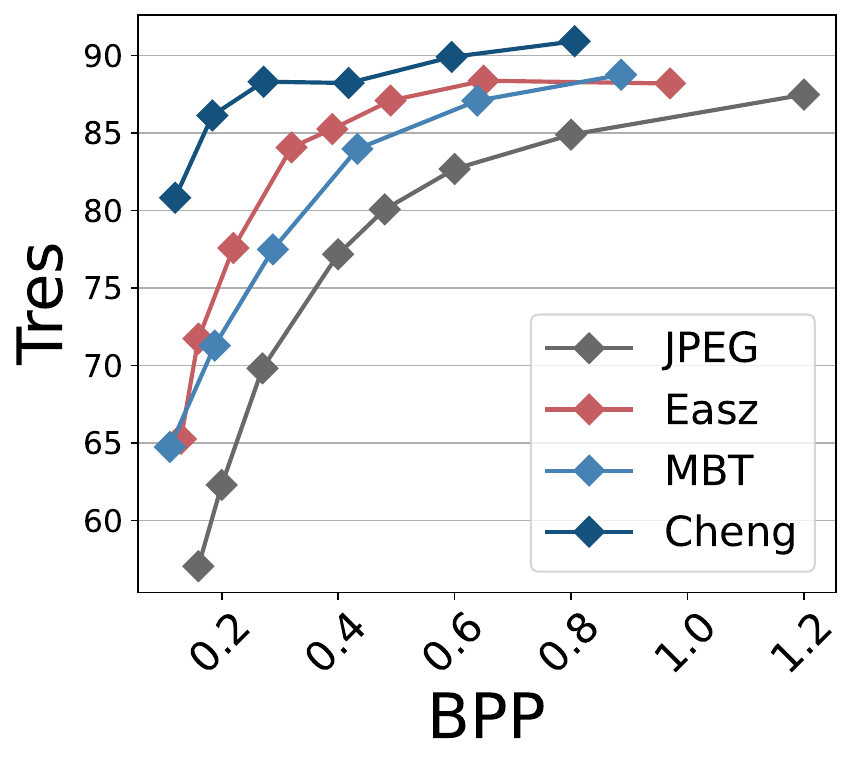}
    \caption{Tres ($\uparrow$ better).}
    \label{fig:distortion_tres}
\end{subfigure}
\begin{subfigure}{0.2\textwidth}
    \includegraphics[width=\textwidth]{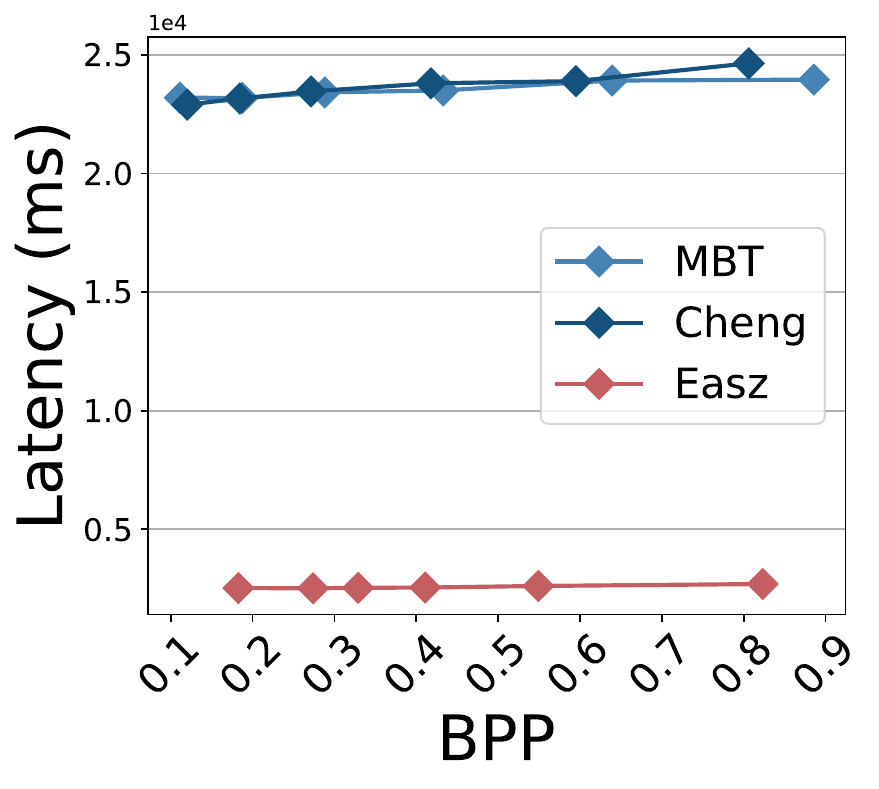}
    \caption{End-to-end Latency.}
    \label{fig:overall_latency}
\end{subfigure}
\caption{Compression performance of \sys, JPEG, MBT and Cheng on three perceptual metrics (a-c). Fig.~\ref{fig:overall_latency} evaluates the end-to-end latency on our testbed.}
\label{fig:ablation}  
\end{figure}

In this experiment, we use JPEG+\sys as the baseline and observe changes in three perceptual metrics at different bitrates (BPP). 
Notably, JPEG alone underperforms compared to two deep-learning compression methods at all compression levels. However, with \sys enhancement, JPEG shows a marked improvement. For the BRISQUE metric specifically, JPEG+\sys exceeds both deep-learning methods. Regarding the Pi metric, JPEG+\sys matches the performance of these methods. With the Tres metric, while JPEG+\sys outdoes MBT, it falls short of Cheng-anchor's results. Overall, \sys boosts JPEG to compete effectively with other state-of-the-art deep-learning compression techniques in each perceptual measure. \sys also outperforms two neural network-based methods in latency, with an average end-to-end latency of 2568ms across different bitrates per pixel, marking an 89\% reduction compared to MBT and Cheng's methods.

%% file: sections/conclusion.tex
\section{Conclusion}

This paper proposes \sys, which overcomes the challenges faced by modern neural image compression on edge devices.  \sys is a comprehensive and efficient data compression framework, which shifts the computational burden to the server side.
 \sys includes an Erase-And-Squeeze process on the edge, coupled with a transformer-based encoder-decoder architecture on the server. Through careful design of the Erase-and-Squeeze strategy, \sys enhances the performance of existing methods while enabling dynamic adjustment of compression levels.
% This is achieved by controlling the erase ratio with only one model on the server. 
As a result, \sys eliminates the computational and storage burdens on the edge while simultaneously boosting compression rates and transmission efficiency.
Our real-world evaluation in an edge-server testbed demonstrates \sys's improvement.
%in compression performance and efficiency, emphasizing its potential for real-world applications.